\documentclass[iop]{emulateapj}

\usepackage{graphicx}
\usepackage{float} 
\usepackage{amsmath}
\usepackage{epstopdf}
\usepackage{epsfig}
\usepackage{color}
\usepackage[breaklinks,colorlinks,citecolor=blue]{hyperref}

\renewcommand{\d}[0]{\mathbf{d}}
\newcommand{\n}[0]{\mathbf{n}}
\newcommand{\dbar}[0]{\bar{\mathbf{d}}}
\newcommand{\Cbar}[0]{\bar{\mathbf{C}}}

\newcommand{\A}[0]{\mathbf{A}}
\newcommand{\s}[0]{\mathbf{s}}
\renewcommand{\a}[0]{\mathbf{a}}

\newcommand{\f}[0]{\mathbf{f}}
\newcommand{\F}[0]{\mathbf{F}}
\newcommand{\B}[0]{\mathbf{B}}
\newcommand{\W}[0]{\mathbf{W}}
\newcommand{\V}[0]{\mathbf{V}}
\newcommand{\T}[0]{\mathbf{T}}
\newcommand{\C}[0]{\mathbf{C}}
\newcommand{\Cell}[0]{C_{\ell}}

\renewcommand{\t}[0]{\mathbf{t}}

\newcommand{\N}[0]{\mathbf{N}}
\newcommand{\M}[0]{\mathbf{M}}
\newcommand{\I}[0]{\mathbf{I}}
\newcommand{\Y}[0]{\mathbf{Y}}
\newcommand{\Yinv}[0]{\mathbf{Y^{-1}}}

\renewcommand{\S}[0]{\mathbf{S}}

\renewcommand{\P}[0]{\mathbf{P}}

\def\WMAP{\textit{WMAP}}

\def\Planck{\textit{Planck}}

\begin{document}

\title{Optimized large-scale CMB likelihood and quadratic \\maximum likelihood power spectrum estimation}

\author{E. Gjerl{\o}w\altaffilmark{1},
  L. P. L. Colombo\altaffilmark{2,3}, H. K. Eriksen\altaffilmark{1},
  K. M. G\'{o}rski\altaffilmark{3}, A. Gruppuso\altaffilmark{4,5},\\
  J. B. Jewell\altaffilmark{3}, S. Plaszczynski\altaffilmark{6} and
  I. K. Wehus\altaffilmark{3}}

\email{eirik.gjerlow@astro.uio.no}

\altaffiltext{1}{Institute of Theoretical Astrophysics, University of
  Oslo, P.O.\ Box 1029 Blindern, N-0315 Oslo, Norway}

\altaffiltext{2}{USC Dana and David Dornsife College of Letters, Arts
  and Sciences, University of Southern California, University Park
  Campus, Los Angeles, CA 90089, USA} 

\altaffiltext{3}{Jet Propulsion Laboratory, California Institute of
  Technology, Pasadena, CA 91109, USA}

\altaffiltext{4}{INAF/IASF Bologna, Via Gobetti 101, Bologna, Italy}

\altaffiltext{5}{INFN, Sezione di Bologna, Via Imero 46, I-40126, Bologna, Italy}

\altaffiltext{6}{Laboratoire de l'Acc\'{e}l\'{e}rateur Lin\'{e}aire,
  Universit\'{e} Paris-Sud 11, CNRS/IN2P3, Orsay, France}

\date{Received - / Accepted -}

\begin{abstract}
  We revisit the problem of exact CMB likelihood and power spectrum
  estimation with the goal of minimizing computational cost through
  linear compression. This idea was originally proposed for CMB
  purposes by Tegmark et al.\ (1997), and here we develop it into a
  fully working computational framework for large-scale polarization
  analysis, adopting \WMAP\ as a worked example. We compare five
  different linear bases (pixel space, harmonic space, noise
  covariance eigenvectors, signal-to-noise covariance eigenvectors and
  signal-plus-noise covariance eigenvectors) in terms of compression
  efficiency, and find that the computationally most efficient basis
  is the signal-to-noise eigenvector basis, which is closely related
  to the Karhunen-Loeve and Principal Component transforms, in
  agreement with previous suggestions. For this basis, the information
  in 6836 unmasked \WMAP\ sky map pixels can be compressed into a
  smaller set of 3102 modes, with a maximum error increase of any
  single multipole of 3.8\% at $\ell\le32$, and a maximum shift in the
  mean values of a joint distribution of an amplitude--tilt model of
  0.006$\sigma$. This compression reduces the computational cost of a
  single likelihood evaluation by a factor of 5, from 38 to 7.5 CPU
  seconds, and it also results in a more robust likelihood by
  implicitly regularizing nearly degenerate modes. Finally, we use the
  same compression framework to formulate a numerically stable and
  computationally efficient variation of the Quadratic Maximum
  Likelihood implementation that requires less than 3 GB of memory and
  2 CPU minutes per iteration for $\ell \le 32$, rendering low-$\ell$
  QML CMB power spectrum analysis fully tractable on a standard
  laptop.
\end{abstract}

\section{Introduction}
\label{sec:introduction}

Through a series of increasingly sensitive experiments measuring the
cosmic microwave background (CMB), led by COBE \citep{mather:1990},
WMAP \citep{bennett:2013} and Planck \citep{planckI:2014}),
cosmologists have during the last two decades established a successful
cosmological concordance model \citep{planckXVI:2014}. According to
this model, the universe is isotropic and homogeneous, and filled with
Gaussian random fluctuations drawn from a nearly scale-invariant
primordial power spectrum; its energy budget is made up by 68\% dark
energy, 27\% dark matter and 5\% baryonic matter. Remarkably, only six
or seven parameters are required to model accurately millions of data
points.

The connection between those millions of data points and the handful
of cosmological parameters is made through the so-called likelihood
function, and cosmological parameter estimation essentially amounts to
mapping out this function, for instance using Markov Chain Monte Carlo
\citep{lewis:2002}, multi-dimensional gridding \citep{mikkelsen:2013}
or non-linear optimization \citep{planckintXVI:2013}. Since the CMB
fluctuations are observed to be (at least close to) Gaussian
distributed, the analytic expression for the likelihood is formally
given by a multivariate Gaussian. However, this expression is of
limited practical use for modern CMB experiments, because of the high
dimensionality of the associated covariance matrix. For Planck, the
number of pixels is $N_{\textrm{pix}}\sim5\times10^{7}$, and since
brute-force likelihood evaluation requires a Cholesky decomposition of
this matrix, computationally scaling as
$\mathcal{O}(N_{\textrm{pix}}^3)$, a single evaluation would cost
$\sim$$10^6$ CPU years and require and require $10^{4}$ TB RAM
\citep[see, e.g.,][for a related discussion]{borrill:1999}.

Obviously, the direct brute-force likelihood approach is not feasible
for modern full-sky CMB experiments, and a few alternative methods
have therefore been proposed and implemented in the literature. These
can largely be broken into two groups. First, the most widely adopted
approach is that of a hybrid likelihood, which simply splits the full
likelihood into two components according to angular scales. Large
angular scales are analyzed using some exact method that fully
accounts for the non-Gaussian nature of the likelihood, whereas small
angular scales are analyzed using faster likelihood approximations
motivated by the Central Value theorem. Usually, the two likelihoods
are sufficiently uncorrelated that they may be joined into a single
all-scale expression either by straight multiplication or by
explicitly accounting for overlap correlations \citep{gjerlow:2013}.
The second group of methods may be characterized as samplers, as for
instance implemented through Gibbs sampling
\citep{jewell:2004,wandelt:2004,eriksen:2004}, which draws samples
from the CMB posterior. The computational scaling of this approach is
$\mathcal{O}(N_{\textrm{pix}}^{3/2})$, and therefore computationally
feasible even for high resolution data. However, further work is
required for this potential to be fully realized, as the computational
expense is still considerable \citep{seljebotn:2014}, and in practice
samplers are still mostly used on large and intermediate angular
scales \citep{planckXII:2014}.

Although one could argue that the ever-advancing progress of computer
technology lessens the need for clever likelihood approximations, one
could at the same time argue that reducing the time needed to perform
likelihood evaluations allows us to expand our field of interest. 
As an example, there is little to gain in terms of computational time
if we restrict our interest to the standard 6-parameter $\Lambda CDM$
model, which can presently be tackled by a standard laptop in a
comfortable time frame. Presently, however, considering extensions
to this model is gaining more and more interest. Such extensions expand
the parameter space of interest, which reintroduces the need to
make our likelihood evaluations as fast as possible while still being
reasonably accurate.

In this paper we revisit the problem of exact brute-force likelihood
evaluation on large angular scales, exploiting the ideas initially
introduced for CMB analysis purposes by \citet{tegmark:1997b} to
reduce the computational cost through linear compression. Rather than
crudely downgrading the data in pixel space until the computational
costs are acceptable, we compress the data into a lower-dimensional
basis set using a more general linear transformation, thereby reducing
computational costs while retaining by far most of the important
information. Further, we also show how this formalism naturally leads
to a very efficient implementation of the Quadratic Maximum Likelihood
(QML) power spectrum estimator.

\section{Basic definitions}
\label{sec:algorithms}

In their most basic form, CMB observations may be
modelled\footnote{Bold lower case letters denote vectors, and bold
  capital letters matrices. In pixel basis, vectors consist of the
  Stokes $I$, $Q$ and $U$ parameters stacked sequentially, and
  matrices consist of $3\times3$ block matrices containing $II$, $IQ$,
  $IU$ etc.} as a linear sum of a cosmological CMB signal, $\s$,
observed by some instrumental beam convolution operator, $\B$, some
set of foreground contaminants, $\f$, and random noise, $\n$,
\begin{equation}
\d = \B\s + \f + \n.
\end{equation}
The signal and noise terms are usually both assumed to be Gaussian
distributed with zero mean and covariances $\S =
\B\left<\s\s^t\right>\B^t$ and $\N = \left<\n\n^t\right>$,
respectively, and the total data covariance matrix is $\C=\S+\N$,
neglecting the foreground term for the moment.

In most cases, the CMB signal is assumed to be isotropic, and it is
therefore particularly convenient to expand this component into
spherical harmonics,
\begin{equation}
\s = \sum_{\ell=0}^{\ell_{\textrm{max}}} \sum_{m=-\ell}^{\ell} s_{\ell
  m} \mathbf{Y}_{\ell m} \equiv \mathbf{Y}\tilde{\s}.
\label{eq:sphharm}
\end{equation}
Here we have defined $\Y$ to be a matrix listing all spherical
harmonics (both spin-0 for temperature and spin-2 for polarization;
see \citet{zaldarriaga:1997} for details) up to some maximum
band limit, $\ell_{\textrm{max}}$ column-wise, and $\tilde{\s}$ to be a
vector containing the spherical harmonics coefficients of $\s$. We
additionally define the symbol $\Yinv$ to denote the inverse spherical
harmonic transform,
\begin{equation}
\tilde{\s} = \int_{4\pi} \Y_{\ell m}^*\s \,d\Omega \equiv \Yinv\s,
\end{equation}
but emphasize that this is not a true inverse of $\Y$, as neither $\Y$
nor $\Yinv$ is square, and any spherical pixelization introduce
non-orthogonality between modes on small angular scales; we only use
$\Yinv$ for high-$\ell$ mode filtering in the combination $\P=\Y\Yinv$
in this paper, for which exact orthogonality is not required.

Under the assumption of statistical isotropy, the signal covariance
takes on a particularly simple form in spherical harmonic space, and
is given by the angular power spectrum, $\Cell$. Assuming further that
the instrumental beam is circularly symmetric and fully described by a
set of Legendre coefficients, $b_\ell$, and that any smoothing effects
from discrete pixelization may be described in terms of an effective
pixel window function, $p_{\ell}$, the harmonic space elements of the
signal covariance matrix reads
\begin{equation}
\tilde{S}_{\ell m, \ell' m'} \equiv b_{\ell}p_{\ell}\left< s_{\ell m} s^*_{\ell' m'}
\right>p_{\ell'}b_{\ell'} = \Cell b_{\ell}^2
p_{\ell}^2\delta_{\ell\ell'}\delta_{mm'}.
\end{equation}
where we have for simplicity defined the power spectrum coefficient,
$C_{\ell}$, to denote a $3\times 3$ block incorporating all
temperature and polarization auto- and cross-spectra,
\begin{equation}
C_{\ell}= \left(\begin{array}{ccc}
C_{\ell}^{TT} & C_{\ell}^{TE} & C_{\ell}^{TB} \\
C_{\ell}^{TE} & C_{\ell}^{EE} & C_{\ell}^{EB} \\
C_{\ell}^{TB} & C_{\ell}^{EB} & C_{\ell}^{BB}
\end{array}\right) 
\end{equation}
From Equation \ref{eq:sphharm} we see that the corresponding signal
covariance matrix in map domain simply reads $\S = \Y\tilde{\S}\Y^T$,
and it is easy to show that the entries of this matrix are given by
the two-point correlation function.

Properly accounting for the foreground term is a far more complicated
problem, and an extensive literature has been written on this topic
\citep[e.g.,][and references therein]{leach:2008,planck2014-a12}. In
this paper we limit ourselves to a very basic foreground model in
which $\f$ may be described by a finite set of spatial templates, each
known perfectly up to an overall amplitude,
\begin{equation}
\f = \sum_i a_i \t_i = \T\a,
\end{equation}
where $\T$ is a matrix listing all templates column-wise, and $\a$ is a
vector of template amplitudes. Accounting for such templates is most
easily implemented by solving the normal equations for $\a$, and
redefining the data vector and data covariance matrix as follows,
\begin{align}
\mathbf{d} &\leftarrow \d - \T\left(\T^t(\S+\N)^{-1}\T\right)^{-1} \T^t
(\S+\N)^{-1}\d \label{eq:temp1} \\ 
\mathbf{N} &\leftarrow \N + \alpha \T\T^t\label{eq:temp2};
\end{align}
here $\alpha$ is a parameter that estimates the uncertainty in the
template fit, and $\alpha \rightarrow \infty$ corresponds to full
projection. However, from a numerical point of view it is more
convenient to set $\alpha$ to a large numerical value, to avoid an
otherwise singular covariance matrix. In this paper, we let $\T$
consist of the monopole and three dipoles, all normalized to a maximum
of unity, and let $\alpha = 10^3$.

With the above data model, the data likelihood depends only on the
angular power spectrum, and is given by a multivariate Gaussian,
\begin{equation}
  \mathcal{L}(\Cell|\d) \equiv P(\d|C_{\ell}) \propto \frac{e^{-\frac{1}{2}\d^T(\S(\Cell)+\N)^{-1}\d}}{\sqrt{|\S(\Cell)+\N|}}
  \label{eq:lnL_basic},
\end{equation}
where we have implicitly accounted for template marginalization by the
redefinitions in Equations \ref{eq:temp1} and \ref{eq:temp2}.  In
principle, this expression can be used directly for CMB power spectrum
or cosmological parameter estimation when coupled to some non-linear
optimization or MCMC implementation. However, as already noted, this
expression contains both a matrix inverse and a determinant, and
therefore scales computationally as
$\mathcal{O}(N_{\textrm{pix}}^3)$. Direct likelihood evaluations are
therefore computationally very expensive, and the main goal of this
paper is to speed up this expression simply by reducing the effective
number of pixels.

\section{The 9-year WMAP low-$\ell$ likelihood}
\label{sec:data}

For pedagogical purposes, we specialize the discussion in this paper
to the low-$\ell$
\WMAP\ likelihood\footnote{http://lambda.gsfc.nasa.gov}, as presented
by \citet{hinshaw:2013}. However, we note that the same approach
should be fully applicable to corresponding \Planck\ low-$\ell$
polarization observations once available.

The 9-year \WMAP\ low-$\ell$ likelihood function is implemented as a
hybrid between a pure temperature likelihood using a
Blackwell-Rao estimator \citep{chu:2005}, and a pure polarization
brute-force likelihood, similar to that described in Equation
\ref{eq:lnL_basic}. Correlations between the two are handled by
explicitly decorrelating the temperature component from the Stokes $Q$
and $U$ maps, given some fixed estimate of the full-sky temperature
sky map and $C_{\ell}^{TE}$ \citep{page:2007}. For computational
speed, the polarization data are degraded onto a very low-resolution
grid, defined by the HEALPix\footnote{http://healpix.jpl.nasa.gov}
pixelization with a resolution parameter of
$N_{\textrm{side}}=8$. This pixelization has a pixel size of
$7^{\circ}\times7^{\circ}$, and supports harmonic modes reliably only
up to $\ell_{\textrm{max}}=16$, although the \WMAP\ likelihood
implementation formally includes modes up to $\ell=23$. After applying
a Galactic mask removing contaminated pixels, a total of 1100
low-resolution polarization ($Q$ and $U$) pixels are included in the
likelihood.

This approach leads to a fast and flexible low-$\ell$
likelihood. However, in the process several assumptions have been
made, most notably that the temperature noise is fully negligible
(enabling the temperature-polarization split), and that the full-sky
temperature modes are well described by the \WMAP\ ILC map. Neither of
these assumptions are obvious (see, e.g., \citealp{finelli:2013} for a
relevant discussion), and in particular the assumption of no
temperature noise has significant consequences in terms of the
effective prior of the likelihood: An absolute mathematical
requirement for any likelihood is that the total covariance matrix,
$\S+\N$, is positive definite, while a softer physical requirement is
that the signal covariance $\S$ alone is positive definite. Enforcing
these requirements consistently is not trivial with a split
likelihood, and we will see in Section \ref{sec:results} that the
\WMAP\ likelihood has both nonphysical ``holes'' as a result of this,
as well as a generally complicated behaviour near the singularity
regions.

A second issue with the \WMAP\ likelihood implementation lies in its
resource requirements. To accelerate likelihood evaluations, the
\WMAP\ code precomputes the Legendre polynomials for each pair of
pixels, and thereby saving CPU time for building the signal covariance
matrix. However, this is costly in terms of memory, and requires 1 GB
RAM already at $N_{\textrm{side}}=8$, which only supports
$\ell\lesssim16$. Doubling the resolution, in order to probe scales up
to $\ell\lesssim32$ increases this requirement to 33 GB, which is more
than most computers can handle comfortably today.

In this paper, we present a more direct implementation of a low-$\ell$
WMAP likelihood that relies only on the brute-force likelihood
expression in Equation \ref{eq:lnL_basic}. Both temperature and
polarization sky maps are considered at a common resolution parameter
of $N_{\textrm{side}}=16$. Otherwise, we adopt data combinations that
are as close as possible to those used for the official WMAP
likelihood \citep{hinshaw:2013}. Specifically, for polarization we
include only the foreground-reduced WMAP Ka, Q and V bands in the
following, not the K-band, which is used for foreground cleaning, or
the W-band, which is known to have worse correlated noise and/or
systematics issues than the other frequencies. For the temperature
component, we adopt the 9-year WMAP ILC map, smoothed to $10^{\circ}$
FWHM.

The individual foreground-reduced polarization frequency maps are
co-added into a single ``clean'' CMB map by inverse noise variance
weighting,
\begin{equation}
\mathbf{d} = \left(\sum_i \N_i^{-1}\right)^{-1}\sum_{i} \N_i \d_i,
\end{equation}
where $\N_i$ is the full covariance matrix for band $i$, which also
take into account the additional noise contribution from the
foreground reduction. The noise covariance of the co-added map reads
\begin{equation}
\N = \left(\sum_i \N_i^{-1}\right)^{-1}.
\end{equation}
We finally add $2\,\mu\textrm{K}$ regularization noise to the ILC
temperature map, to make the temperature covariance matrix
invertible. The full noise covariance thus consists of a diagonal
temperature block and a dense polarization block, with no cross-terms
between the two. 

We adopt the WMAP KQ85 mask for the temperature component, and the P06
mask for the polarization components \citep{bennett:2013}, leaving a
total of 2326 temperature and 4510 polarization ($Q$ and $U$) pixels
for analysis, or a total of 6836 elements in the data vector. The
instrumental beams are taken to be a perfect Gaussian of $10^{\circ}$
FWHM for the temperature, and a Gaussian of 30.6 arcmin for
polarization, corresponding roughly to the Q-band beam, and adopted as
a rough average of the three channels; its impact is very small for
the multipoles considered in the following, having a minimum amplitude
of $b_{\ell}=0.993$ at $\ell=30$.

\section{Linear compression and basis definitions}
\label{sec:compression}

\subsection{Basic formalism}

Any linear transformation of a set of Gaussian random variables
results in another set of Gaussian random variables. Consider
therefore some linear combination on the form
\begin{equation}
\bar{\d} = \P\d,
\end{equation}
where $\P$ is some $N \times N_{\textrm{pix}}$ transformation matrix
with $N \le N_{\textrm{pix}}$, and $\bar{\d}$ is a transformed data
vector. If $\d$ is a zero-mean Gaussian field with covariance $\C$,
then $\bar{\d}$ will be a zero-mean Gaussian field with covariance
$\bar{\C} = \P\C\P^t$. With the data model described above, the
corresponding likelihood for these compressed data therefore reads
\begin{align}
\nonumber
  \mathcal{L}(\Cell|\bar{\d}) &\propto
  \frac{e^{-\frac{1}{2}\bar{\d}^T(\bar{\S}(\Cell)+\bar{\N})^{-1}\bar{\d}}}{\sqrt{|\bar{\S}(\Cell)+\bar{\N}|}}\\
  &= \frac{e^{-\frac{1}{2}\P^T\d^T(\P\S(\Cell)\P^T+\P\N\P^T)^{-1}\P\d}}{\sqrt{|\P\S(\Cell)\P^T+\P\N\P^T|}}
  \label{eq:lnL_compressed}.
\end{align}
The interesting question is now whether there exists some
transformation $\P$ that retains the relevant information in $\d$ with
a smaller number of data points, $N < N_{\textrm{pix}}$.

\begin{figure}[t]
\begin{center}
\mbox{\epsfig{figure=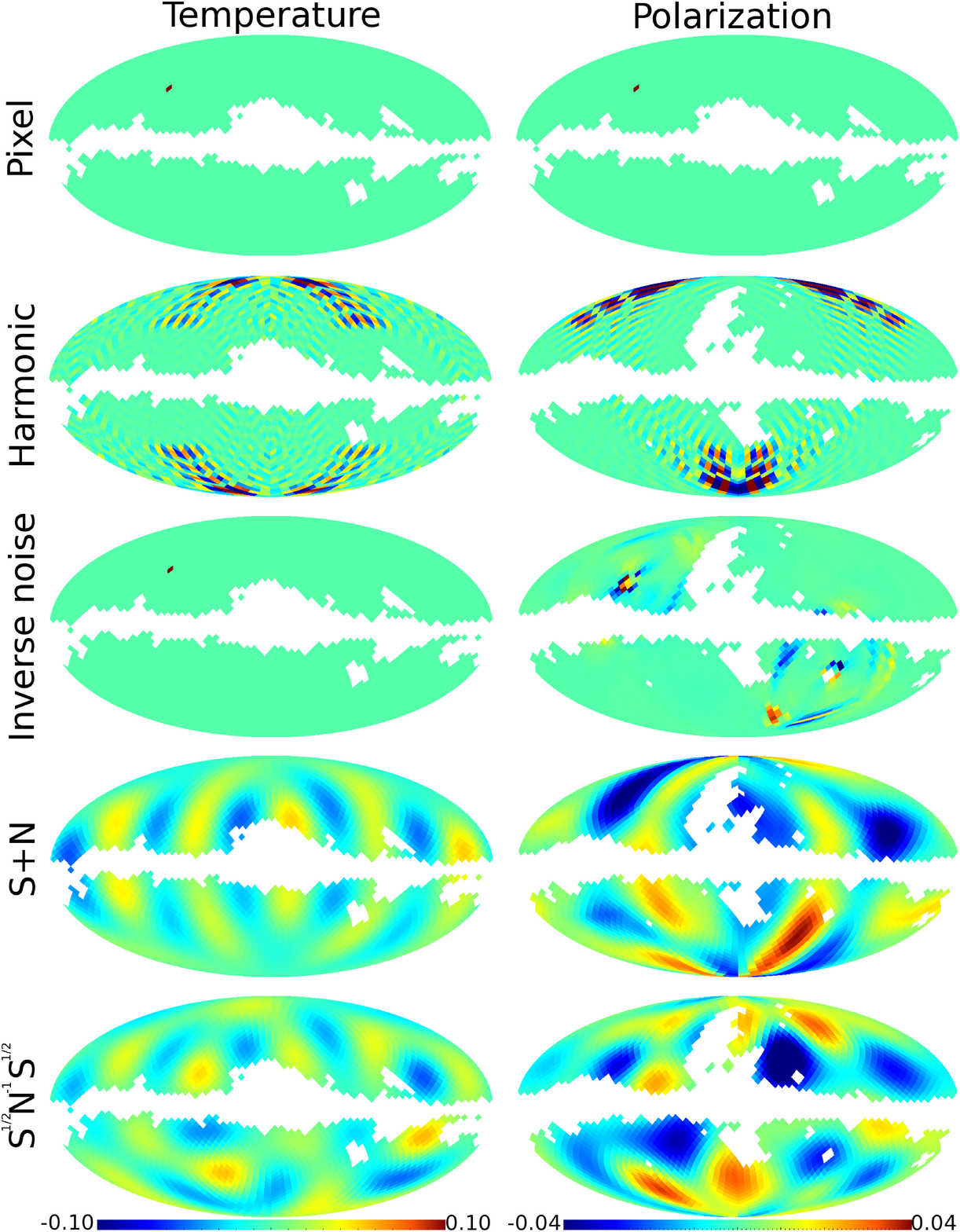,width=\linewidth,clip=}}
\end{center}
\caption{Example basis vectors for temperature (\emph{left column})
  and polarization (\emph{right column}) for each of the five basis
  sets considered in this paper, computed from the 9-year \WMAP
  data. In each case, the basis vector with the 30th highest
  eigenvalue is shown, and only the Stokes $Q$ component is shown for
  polarization; the Stokes $U$ components look qualitatively similar.}
\label{fig:basis}
\end{figure}

\begin{figure}[t]
  \begin{center}
    \epsfig{figure=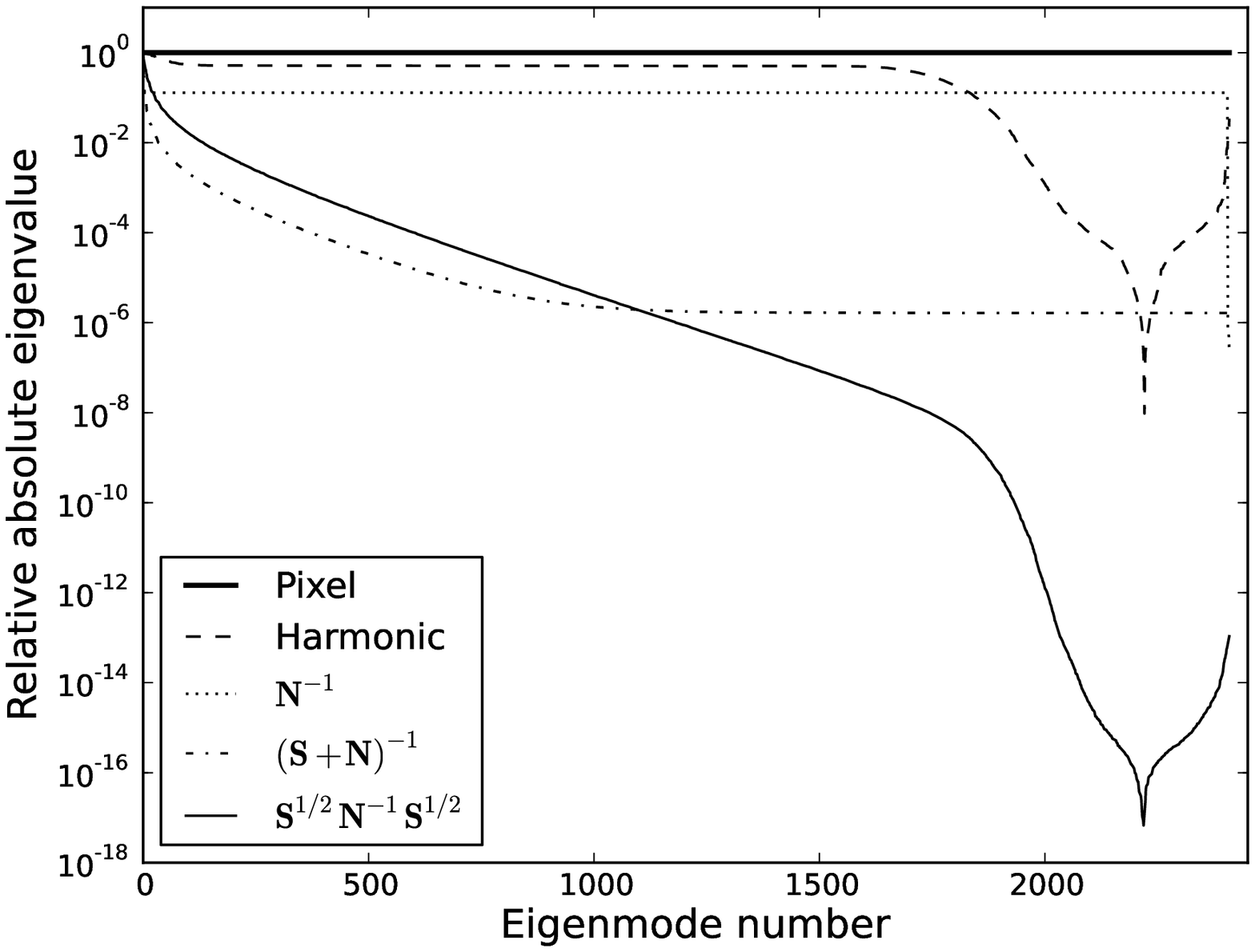, width=0.9\linewidth}
    \epsfig{figure=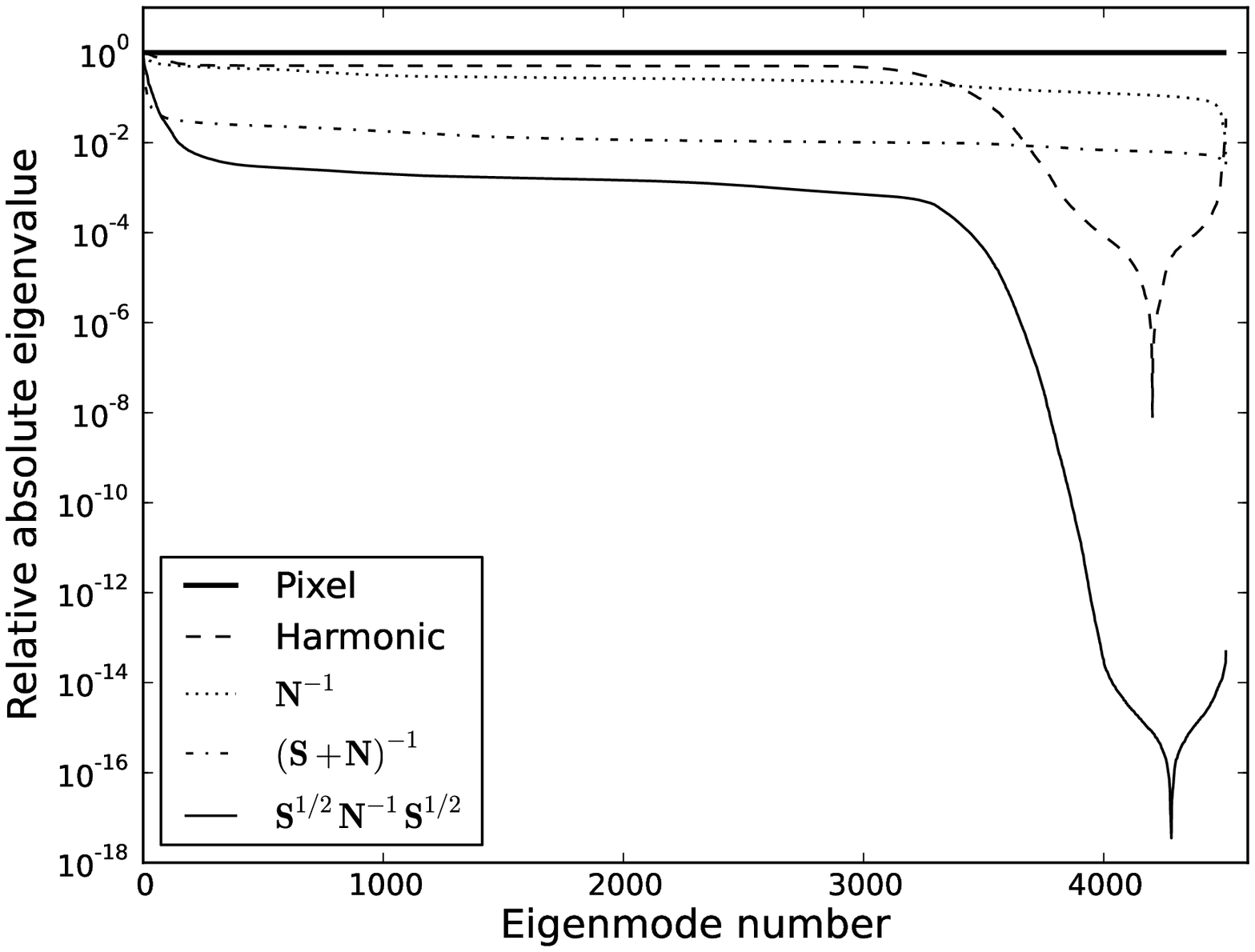, width=0.9\linewidth}
  \end{center}
  \caption{Eigenvalue spectra for all five bases defined in the text,
    shown both for temperature (\emph{top}) and polarization
    (\emph{bottom}), as evaluated for the 9-year \WMAP\ data, using
    no high-$\ell$ truncation. The
    eigenvalues are normalized to the maximum value, and sorted
    according to decreasing values; increasing values from left to
    right indicate negative eigenvalues. }
  \label{fig:eigspecs}
\end{figure}

Before explicitly defining a set of candidate bases, it is useful to
introduce some additional notation. First, since there are always
parts of the sky that are unavailable for cosmological CMB analysis
due to foreground contamination from our own Galaxy, we introduce a
pixel space masking operator, $\M$, defined in terms of an
$N_{\textrm{mask}} \times N_{\textrm{pix }}$ matrix that contains one
row for each unmasked pixel, with the value 1 in the column
corresponding to the pixel number; all other entries are zero. When
applied to a full-sky data vectors, this operator simply picks out the
unmasked pixels, leaving all values numerically unchanged.

Second, we define a harmonic space truncation operator, $\P_h \equiv
\M \Y_{\ell_{\textrm{t}}}\Yinv_{\ell_\textrm{t}}\M^T$, where only
spherical harmonics up to some truncation multipole,
$\ell_{\textrm{t}} \le \ell_{\textrm{max}}$, are included in the
spherical harmonics operator. This operator filters out any spherical
harmonics above $\ell_{\textrm{t}}$, evaluated only over masked
pixels; since only masked pixels are included, the operator is not a
sharp operator in multipole space, but rather corresponds to a
pseudo-$a_{\ell m}$ projection operator with non-zero coupling to
multipoles above $\ell_{\textrm{t}}$ \citep[e.g.,][]{hivon:2002}.

Third, we define $[\A]_\epsilon$ as the set of eigenvectors of $\A$
with a fractional eigenvalue larger than $\epsilon$ relative to the
maximum eigenvalue. That is, let $\V$ be the matrix containing the
eigenvectors of $\A$, and $\W$ be the diagonal matrix of eigenvalues,
such that $\A = \V\W\V^t$; then $[\A]_\epsilon$ contains all columns
of $\V$ with an eigenvalue larger than $\epsilon\cdot \max(\W)$. This
operator removes modes with low eigenvalues, which in turn will be
used to eliminate modes with low signal-to-noise ratio. However,
because of the very different signal amplitudes in temperature and
polarization, we define two different eigenvalue thresholds,
$\epsilon_\textrm{T}$ and $\epsilon_{\textrm{P}}$, for temperature or
polarization modes, and set the temperature-polarization
cross-elements in $A$ to zero before performing the eigenvalue
decomposition.

\subsection{Basis sets}

With the above notation, we define five candidate bases to be
considered for further analysis,
\begin{equation*}
\begin{array}{lll}
\P_1 &= \M & \textrm{Pixel} \\
\P_2 &= [\P_h]_\epsilon\M & \textrm{Harmonic} \\
\P_3 &= [\P_h\N^{-1}\P_h^t]_\epsilon\M & \textrm{Inverse noise} \\
\P_4 &= [\P_h(\S+\N)\P_h^t]_\epsilon\M & \textrm{Signal-plus-noise} \\
\P_5 &= [\P_h(\S^{1/2}\N^{-1}\S^{1/2})\P_h^t]_\epsilon\M &
\textrm{Signal-to-noise},
\end{array}
\end{equation*}
where $\S = \S(C_{\ell}^{\textrm{fid}})$ is the signal covariance
matrix computed from some fiducial model\footnote{We set
  $C_{\ell}^{BB,\textrm{fid}}=C_{\ell}^{EE,\textrm{fid}}$ when
  constructing the basis signal covariance matrix in the following
  analyses, to ensure good sampling of both spectra spectra.}.  Each
basis is either commonly encountered in the literature (i.e., pixels,
harmonics, signal-to-noise eigenmodes) or have a well-defined specific
purpose (e.g., the inverse noise basis is particularly well suited to
test systematics by suppressing poorly measured modes, while the
signal-plus-noise basis corresponds to numerical regularization of the
data covariance matrix). It is also worth noting that the
signal-to-noise basis is closely related to the Karhunen-Loeve (or
Principal Component) transform originally proposed for cosmological
applications by \citet{tegmark:1997b}. Potential dependence on the
assumed fiducial spectrum, $C_{\ell}^{\textrm{fid}}$, is considered in
Section \ref{sec:results}; we find no significant detrimental effects
by adopting a power spectrum far from the best-fit spectrum.

There are two tunable parameters in this framework,
$\ell_{\textrm{t}}$ and $\epsilon$, both of which have a very
intuitive interpretation: Lowering $\ell_{\textrm{t}}$ removes
high-$\ell$ spherical harmonic modes, while increasing $\epsilon$
removes low signal-to-noise modes. However, it is important to note
that no choice of either $\ell_{\textrm{t}}$ or $\epsilon$ can ever
\emph{bias} the power spectrum, but only modify the
uncertainties. Linear compression simply amounts to removing
irrelevant modes, and is mathematically fully equivalent to removing
masked pixels. However, lowering $\ell_{\textrm{max}}$ (as opposed to
$\ell_{\textrm{t}}$) will both bias the power spectrum and increase
the $\chi^2$, because it changes the data model, not simply the data
selection. This is an important difference between our approach and
that implemented by the official WMAP polarization likelihood code,
which simply downgrades the actual sky maps from
$N_{\textrm{side}}=16$ to 8.

Before proceeding with basis optimization, it is useful to build some
intuition about the various basis candidates. In Figure
\ref{fig:basis} we therefore show an example basis vector, and in
Figure \ref{fig:eigspecs} we show the eigenvalue spectrum, for each
basis as computed from the 9-year \WMAP\ data (Section
\ref{sec:data}). The example basis vectors all correspond to the
vector with the 30th largest eigenvalue, $\epsilon_{30}$, for both
temperature and polarization. Only the Stokes $Q$ field is shown for
polarization, as Stokes $U$ looks qualitatively similar.

Starting with the pixel basis, we see in Figure \ref{fig:basis} that
each pixel corresponds in this case to an independent basis
vector. Furthermore, as seen in Figure \ref{fig:eigspecs}, the
eigenspectrum is completely flat, and no truncation limit, $\epsilon$,
can remove any degrees of freedom. The pixel basis is therefore always
complete, and all information stored in the uncompressed data is (by
definition) retained with this basis. In the following, we adopt the
pixel basis as the reference against which we measure data loss for
other bases.

The second row in Figure \ref{fig:basis} shows a temperature and
polarization mode of the spherical harmonics basis, and the dashed
line in Figure \ref{fig:eigspecs} shows its eigenspectrum. Both of
these highlight a problematic feature with this particular basis: It
is susceptible to numerical errors at high multipoles. Ideally,
$\Y_{\ell_{\textrm{t}}}\Yinv_{\ell_\textrm{t}}$ should be identically
equal to one for $\ell \le \ell_{\textrm{t}}$ and zero
otherwise. However, because all operations are done on a finite
pixelization that supports only a finite number of multipoles, there
is always some leakage between multipoles in this
operator. Furthermore, one is not guaranteed that $\int Y_{\ell m}
Y_{\ell m}^* d\Omega$ will be smaller than one. On the contrary, the
worst-behaved modes often have a square-integral substantially larger
than one. This is observed as three distinct regions in the
eigenspectrum of the spherical harmonics basis: The flat plateau from
about 50 to 1500 corresponds to well resolved modes with good support
on the masked sky; the rapid decrease above 1500 corresponds to modes
that are filtered either by the high-$\ell$ truncation operator or are
degenerate because of the sky mask\footnote{Note that the absolute
  value of the eigenvalue is plotted here; the sharp feature indicates
  the mode for which the eigenvalue becomes negative.}, and, finally,
the modes below 50 are numerically unstable high-$\ell$ modes that
have an eigenvalue larger than 1. The harmonic mode shown in Figure
\ref{fig:basis} is an example of such a mode.

The third basis corresponds to the eigenvectors of the inverse noise
covariance matrix. For the low-resolution \WMAP\ data, this matrix is
given by a spatially constant regularization noise RMS amplitude of
$2\mu\textrm{K}$ for temperature, and the actually measured
instrumental noise covariance for polarization, including both
scanning strategy and correlated noise effects. For temperature, the
inverse noise basis functions are therefore identical to the pixel
basis, with one pixel per value, with one exception: This basis
explicitly highlights the effect of foreground template projection in
the form of a sharp drop in the eigenspectrum, corresponding to the
monopole and dipoles modes that are manually assigned a numerically
large uncertainty. For polarization, the dominant feature is the
scanning strategy, which is clearly seen in the example basis mode in
Figure \ref{fig:basis}. This basis may be useful for systematics
studies, since instrumental systematics are often strongly associated
with poorly measured modes.

The fourth basis is defined as the eigenvectors of the total data
covariance matrix, $\S+\N$. This could be a relevant basis for cases
that have an ill conditioned covariance matrix, as for instance often
happens for strongly signal-dominated temperature data, $\S >>
\N$. Since $\S$ by construction is spanned by
$(\ell_{\textrm{max}}+1)^2 < N_{\textrm{pix}}$ modes, this situation
leads to a poorly conditioned total covariance matrix that needs to be
regularized before further analysis. The two most common approaches
are either to add a small amount of white noise to the data (known as
``regularization noise'') or to increase $\ell_{\textrm{max}}$ beyond
the Nyquist limit formally supported by the pixelization. A third
option would be to use the $\S+\N$ basis proposed here, which simply
removes by hand poorly conditioned modes from the data set. 

Finally, the fifth basis is given by the eigenvectors of the
signal-to-noise covariance matrix, $\S^{1/2}\N^{-1}\S^{1/2}$, written
in an explicitly symmetric form to minimize numerical errors. In this
case, a prior spectrum is introduced that allows one to select modes
based on individual signal-to-noise ratio, only retaining those that
actually contribute with useful information. Several variations of
this has already been discussed extensively in the literature,
resulting in various implementations of the same underlying ideas, two
of which are the Karhunen-Loeve and Principal Component
transforms. This was also the basis set originally proposed by
\citet{tegmark:1997a}. For a related application to non-Gaussianity,
see \citet{rocha:2001}.

\subsection{A condition number based prior}
\label{sec:regprior}

\begin{figure}[t]
  \begin{center}
    \epsfig{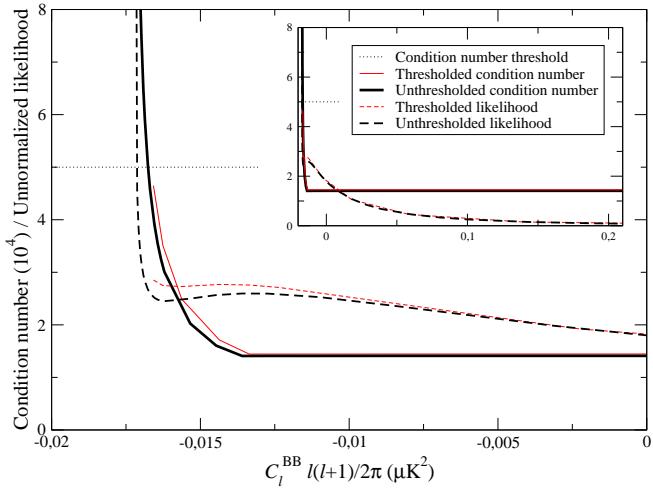}
  \end{center}
  \caption{Regularizing the likelihood with a condition number
    prior. Dashed curves show slices through
    $\mathcal{L}(C_{2}^{BB})$ while fixing all other multipoles at
    their maximum-likelihood points. Solid lines show the condition
    number of the data covariance matrix, $\S+\N$, as a function of
    power spectrum. Thick curves show results with no prior on the
    condition number, and thin curves show results when requiring the
    condition number to be smaller than 50\,000. Condition number
    regularization eliminates likelihood artifacts near the
    singularity boundary.}
  \label{fig:lnL}
\end{figure}

Before proceeding to further analysis, we make a minor comment on a
technical issue already mentioned in Section \ref{sec:data}, namely
that the total data covariance matrix, $\C$ must be positive definite
in order for the likelihood to be well defined. Intuitively and not
mathematically rigorously, this condition breaks down whenever
$S_{\ell m, \ell m} < -N_{\ell m, \ell m}$. However, the likelihood
surface actually become unstable well before this limit because of
numerical errors, as illustrated in Figure \ref{fig:lnL}. The only
difference between the main frame and the inset is the $x$-axis range.
First, the dashed thick line shows an (arbitrarily normalized) slice
through our pixel basis likelihood for $\mathcal{L}(C_{2}^{BB})$,
keeping all other multipoles fixed at their maximum-likelihood
values. This slice exhibits perfectly normal behaviour for large
values of $C_{2}^{BB}$, following roughly the behaviour of an inverse
Gamma distribution. However, near the value of
$C_{2}^{BB}=-0.0175\mu\textrm{K}^2$ the likelihood rapidly increases,
and essentially diverges to infinity.

This behaviour is a generic feature of any likelihood near the
boundary at which it becomes singular: Even if the matrix may be
positive definite and invertible, the numerical value cannot be
trusted sufficiently near the singularity boundary. Fortunately, this
problem can be resolved in several ways, and our preferred solution is
by monitoring the covariance matrix condition number, i.e., the ratio
of the largest to smallest eigenvalue. This quantity is shown as a
solid thick line in Figure \ref{fig:lnL} for the above case. For any
power spectrum value not close to the singularity boundary, we see
that the condition number is highly stable, with a numerical value
around 15\,000 for this particular case. However, once the covariance
matrix approaches singularity, it starts to increase rapidly, and it
does so sooner than the actual likelihood. The combination of a high
degree of stability within the main parameter volume, and rapid
increase toward the edges makes the condition number an effective
monitor of the likelihood robustness. Therefore, rather than requiring
that the covariance matrix simply must be positive definite (as for
instance enforced by the official \WMAP\ likelihood), we demand that
the condition number must be smaller than some pre-defined
threshold. The specific value of this threshold must be determined by
some initial likelihood scans, but in practice this is very
straightforward. For the above basis, we adopt a numerical threshold
of 50\,000, and the resulting regularized likelihood is shown as a
dashed thin line. The slight difference with respect to the
unregularized likelihood at higher values is due to the slightly
different maximum-likelihood power spectrum coefficients at other
multipoles caused by the same prior.

\section{Efficient and stable QML implementation}
\label{sec:qml}

The formalism described in Section \ref{sec:compression} can be used
to derive a computationally efficient variation of the Quadratic
Maximum Likelihood (QML) estimator, initially introduced by
\citet{tegmark:1997a} and \citet{bond:1998} as an efficient route to
the maximum-likelihood CMB power spectrum. For example applications,
see, e.g., \citet{gruppuso:2009,gruppuso:2011,gruppuso:2013} and
references therein. Let $\C_{,b} = \partial\C/\partial C_{b}$ denote
the derivative of the data covariance matrix with respect to some
power spectrum parameter, $C_b$, with $b =
\{\ell_1,\ldots,\ell_{n}\}$, where $n$ denotes the number of
multipoles included in the spectral bin. The first derivative and
Fisher matrix of the log-likelihood may then be written as
\begin{align}
\frac{\partial\ln\mathcal{L}}{\partial C_{b}} &= \frac{1}{2}
\textrm{tr}\left[(\dbar\dbar^t-\Cbar)(\Cbar^{-1}\Cbar_{,b}\Cbar^{-1}) \right]
\label{eq:deriv}
\\ 
F_{bb'} &= \frac{1}{2}
\textrm{tr}\left[\Cbar^{-1}\Cbar_{,b}\Cbar^{-1}\Cbar_{,b'} \right];
\label{eq:fisher}
\end{align}
see Section IIC of \citet{bond:1998} for full details.

The QML estimator is now defined as follows:
\begin{enumerate}
\item Make some initial guess at the power spectrum, $C_{b}^{(0)}$
\item Update the spectrum according to the following rule,
\begin{equation}
C_{b}^{(i)} = C_{b}^{(i-1)} + \sum_{b'} (\F^{-1})_{bb'}
\frac{\partial\ln\mathcal{L}}{\partial C_{b'}}
\label{eq:QML_unstable}
\end{equation}
\item Iterate until convergence
\end{enumerate}
This algorithm is closely related to the Newton-Raphson optimization
method, with the one difference that it employs the (computationally
cheaper) Fisher matrix instead of the curvature matrix. The two
algorithms converge to the same (maximum-likelihood) solution
\citep{bond:1998}.

In this paper, we note that Equations \ref{eq:deriv} and
\ref{eq:fisher} can be slightly rewritten to facilitate fast numerical
evaluation. Specifically, the signal matrix may be written as
$\bar{\C} = \P\Y\tilde{\S}\Y^{\dagger}\P^t +\bar{\N}$, where
$\tilde{\S}$ is the \emph{full-sky} signal covariance matrix in
harmonic space, and all geometry and data selection effects are
encoded in the constant projection operators $\P$ and $\Y$. The
derivative of this matrix with respect to $C_{b}$ reads
\begin{equation}
\frac{\partial \C}{\partial C_b} = \P\Y\I_b\Y^{\dagger}\P^t,
\end{equation}
where $\I_b$ is a harmonic space matrix containing the value 1 for
entries containing $C_{\ell}$ in $\tilde{\S}$ for $\ell \in b$, and
otherwise 0; it is very sparse, and multiplication with this matrix is
fast.

Inserting this expression into Equation \ref{eq:deriv} and
\ref{eq:fisher}, and noting that the trace operator is invariant under
cyclic permutations, we see that 
\begin{align}
\frac{\partial\ln\mathcal{L}}{\partial C_{b}} &= \frac{1}{2}
\textrm{tr}\left[(\Y^{\dagger}\P^t\Cbar^{-1})(\dbar\dbar^t-\Cbar)(\Cbar^{-1}\P\Y)\I_b \right]
\\ 
F_{bb'} &= \frac{1}{2}
\textrm{tr}\left[(\Y^{\dagger}\P^t\Cbar^{-1}\P\Y)\I_b(\Y^{\dagger}\P^t\Cbar^{-1}\P\Y)\I_{b'} \right].
\end{align}
While these expressions look somewhat formidable at first sight, they
are in fact computationally very efficient. Starting with the first
derivative, the important point is that all multipole dependencies
have been factorized away from expensive dense matrix products. After
precomputing $\P\Y$ (which only has to be done once for every basis
set) and grouping the matrix products as indicated with parentheses in
the above equations, the computational cost of the first derivative is
given by only two matrix products plus one Cholesky
factorization/solve, and the total memory consumption is equivalent to
four dense matrices. The memory consumption is independent of the
number of power spectrum bins, and the CPU time is only weakly
dependent on the number of bins, involving only a single sparse trace
evaluation.

A similar consideration holds for the Fisher matrix. In this case, the
main computational cost lies in evaluating
$\Y^{\dagger}\P^t\Cbar^{-1}\P\Y$ once, at the cost of one Cholesky
factorization/solve and one matrix multiplication. Computing the
remaining product and traces is computationally fast, because of the
high sparsity of the $\I_b$ operator.

%\subsection{Stabilized QML estimation}

The iterative QML algorithm as described above has one major weakness:
The power spectrum proposed in iteration $i$ does not necessarily
yield a positive definite total data covariance matrix,
$\bar{\C}$. This typically happens whenever one or more likelihood
conditionals have a sharp edge beyond which (symbolically) $S_{\ell m}
< -N_{\ell m}$, which is not uncommon in the noise dominated regime
(see Section \ref{sec:results} for explicit examples).

As a safe-guard again this problem, we modify the QML algorithm as
follows:
\begin{enumerate}
\item Make some initial guess at the power spectrum, $C_{b}^{(0)}$
\item Update the spectrum according to the following rule,
\begin{equation}
C_{b}^{(i)} = C_{b}^{(i-1)} + \alpha \sum_{b'} (\F^{-1})_{bb'}
\frac{\partial\ln\mathcal{L}}{\partial C_{b'}},
\label{eq:QML}
\end{equation}
where the step length, $\alpha$, maximizes
$\mathcal{L}(C_{b}^{(i)})$. We implement the latter optimization with
a standard line optimizer (\texttt{linmin}; Press et al.\ 2007).
\item Convergence is defined when the log-likelihood has changed by
  less than 0.1 over the last three iterations
%\item In exceptional cases, one many find $|\alpha| < 0.01$, in which
%  case one or more highly non-Gaussian conditionals prevent efficient
%  joint convergence. In these cases, we optimize each conditional
%  independently. 
\end{enumerate}
The underlying intuition is simply to tune the step size along the
proposed QML direction such that the likelihood is maximized. Each
step will necessarily lead to a higher likelihood value, and the
algorithm cannot diverge. 

Unfortunately, this stability comes at a non-negligible computational
cost, as one now has to perform a non-linear optimization within each
main QML iteration, and this operation requires repeated likelihood
evaluations. However, since each likelihood evaluation is quite fast
due to the compression step described above (after all, the likelihood
function is designed to be an active component in an MCMC cosmological
parameter estimation framework), this is not a showstopper; the
benefit of additional stability more than compensates for this
expense.

Before turning to applications, we make one note regarding error
estimation. Often, $\sqrt{(\F^{-1})_{bb}}$ is adopted as an
uncertainty on the QML estimate, a choice that is primarily driven by
computational efficiency. In this paper, we quote asymmetric 68\%
confidence limits, computed by mapping out the likelihood
conditionally around the maximum likelihood point for each parameter,
and finding the smallest range that encompass 68\% of the conditional
likelihood volume.

\section{Basis optimization}
\label{sec:optimization}

\begin{figure}[t]
  \begin{center}
    \epsfig{figure=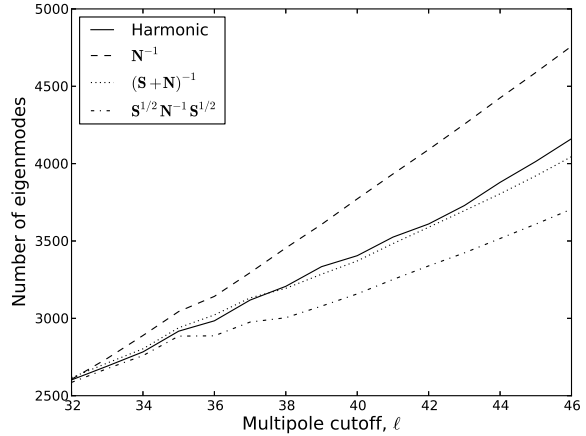, width=0.9\linewidth}
  \end{center}
  \caption{Number of modes required for the Fisher uncertainty to
    increase by no more than 10\% relative to the pixel basis,
    including modes between $2\le\ell\le32$ and plotted as a function
    of truncation multipole.}
  \label{fig:Nmodes_bases_comp}
\end{figure}

We now turn our attention to basis set optimization, considering each
of the five candidates defined in Section \ref{sec:compression} as
applied to the 9-year WMAP data described in Section
\ref{sec:data}. In our framework, basis optimization corresponds
simply to determining the harmonic space truncation multipole,
$\ell_{\textrm{t}}$, and eigenvalue thresholds, $\epsilon_\textrm{T}$
and $\epsilon_\textrm{P}$, that result in the smallest number of
accepted modes under the constraint that the information content over
some range of multipoles is conserved. For the main analysis, we
consider $2\le\ell\le32$ to be the multipole range of interest,
matching that of the official \WMAP\ temperature likelihood, and as a
secondary test, we consider a case in which the polarization range is
constrained to $\ell\le10$, directly targeting the low-$\ell$ $EE$
reionization peak.

\begin{figure}[t]
  \begin{center}
    \epsfig{figure=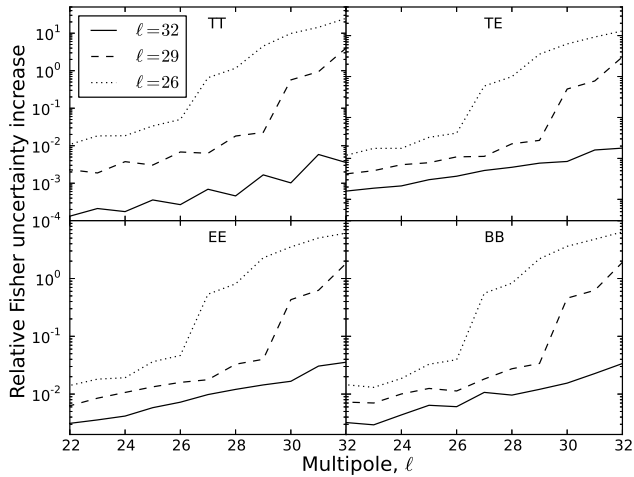, width=1.1\linewidth}
  \end{center}
  \caption{Relative Fisher uncertainty increase as a function of
    truncation multipole for each of the four cosmologically
    interesting power spectra ($C_{\ell}^{TT}$, $C_{\ell}^{TE}$,
    $C_{\ell}^{EE}$ and $C_{\ell}^{BB}$), evaluated for the
    signal-to-noise basis.}
  \label{fig:reldiff}
\end{figure}

The goal of our first test is to compare the efficiency of the five
candidate bases. For this, we base our statistic on the Fisher
information: For each combination of truncation multipole and
eigenvalue thresholds, we compare Fisher uncertainty (i.e.,
$\F_{ii}^{-1/2}$) for the proposed basis with the corresponding value
computed from the full pixel basis. We then require that the
uncertainty must not increase by more than 10\% for any multipole
within the range of interest. Thus, this test is designed only to
compare the relative compression efficiency of the various bases, not
measure absolute information content, as correlations between
multipoles are not properly quantified.

% Eirik's original text -- should be rewritten to fit the text below
%
%It is noteworthy that the number of modes needed is an increasing function of
%the trunctation multipole in all cases. Instinctively, this might seem odd,
%since we're only testing up to $\ell=30$ in all cases, and one would think that
%the number of modes needed should be fixed (as long as we have convergence).
%And it may actually be the case that no more than this is needed for any
%truncation. However, our mode cutoff criterion is based on eigenvalue power
%only, not on how much they contribute to the Fisher information. Thus,
%truncating at higher multipoles means admitting more (non-informative) modes
%into the eigenvalue calculation, and if those modes happen to have high
%eigenvalues, they will be admitted in the new basis. Therefore, in order to 
%include the informative modes, we must admit more modes than for lower
%truncation multipoles.

The results from these calculations are summarized in Figure
\ref{fig:Nmodes_bases_comp}, plotting the lowest number of accepted
modes for each basis as a function of truncation multipole. First, we
see that higher truncation multipoles generally require more modes in
the basis in order to produce stable low-$\ell$ results. This makes
sense because many of the newly added high-$\ell$ modes have a higher
eigenvalue than the some of the previous low-$\ell$ modes, and
therefore more modes have to be included to retain the same low-$\ell$
information. Selecting modes based on harmonic content rather than
eigenvalue would circumvent this issue. Second, and this is the main
point of the plot, we see that the four candidate bases behave
quantitatively differently: While the inverse noise basis require more
than 4500 modes to produce robust results for high truncation
multipoles, only 3500 modes are needed in the signal-to-noise basis,
corresponding to a reduction of 22\% in number of modes or a
theoretical speed-up of 2. The two other bases lie in between, and are
fairly close to each other. All four candidate bases achieve
substantial compression compared to the original pixel basis including
a total of 6836 modes. In the following, we adopt the signal-to-noise
basis as our default compression basis.

In Figure \ref{fig:reldiff} we plot the relative Fisher uncertainty
increase as a function of truncation multipole for each of the four
cosmologically interesting power spectra ($C_{\ell}^{TT}$,
$C_{\ell}^{TE}$, $C_{\ell}^{EE}$ and $C_{\ell}^{BB}$) for this
basis. Decreasing $\ell_{\textrm{t}}$ gradually from 32 to 26, we
observe two main effects. First, the most striking feature is that the
uncertainties increase by almost an order of magnitude for any
multipoles at $\ell > \ell_{\textrm{t}}$. However, they do not become
infinite, because of the non-orthogonality introduced by the mask. In
other words, there is information about high multipoles in cut-sky
harmonics (``pseudo-$a_{\ell m}$s''). Conversely, the second effect is
that the uncertainty also on multipoles below $\ell_{\textrm{t}}$
increase when removing high-$\ell$ modes, gradually increasing the
low-$\ell$ noise floor.

\begin{figure}[t]
  \begin{center}
    \epsfig{figure=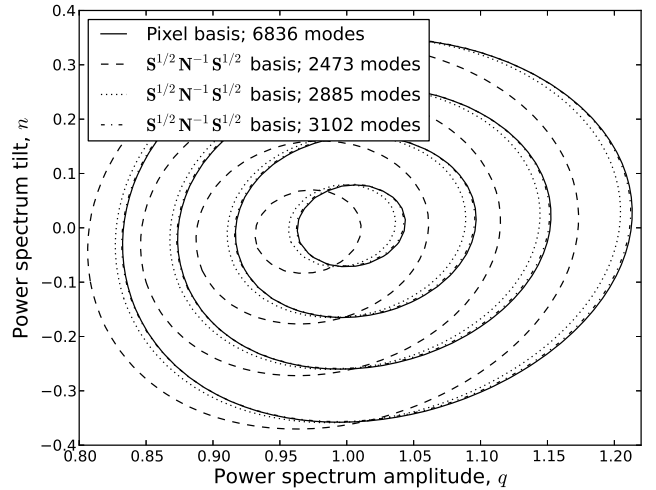, width=\linewidth}
  \end{center}
  \caption{Two-parameter amplitude--tilt likelihoods for various
    eigenmode cutoffs for the signal-to-noise basis (broken contours),
    compared to the corresponding likelihood evaluated from the pixel
    basis (solid contours). The contours indicate the peak and 1, 2
    and $3\sigma$ confidence regions, respectively.}
  \label{fig:twopar}
\end{figure}

From Figure \ref{fig:Nmodes_bases_comp} we know that no Fisher uncertainties
between $2\le\ell\le32$ increase by more than 10\% when including 2500
modes or more in the signal-to-noise basis. However, this is a quite
crude criterion, and not a sufficient criterion for establishing a
proper production likelihood; for this we have to make sure that
correlations are also properly accounted for. We therefore define a
more directly applicable statistic through a simple two-parameter
amplitude--tilt model on the form $C_{\ell}(q,n) =
q\left(\ell/\ell_{\textrm{pivot}}\right)^n C_{\ell}^{\textrm{fid}}$,
and map out the 2D $(q,n)$ likelihood for each effective basis. The
search is done in terms of number of modes, and only the
signal-to-noise basis is subjected to this analysis.

A subset of the results derived in this calculation is shown in Figure
\ref{fig:twopar}, in the form of 2D likelihood contours. Here we see
that when including only 2473 modes, which resulted in less than 10\%
increase in any single multipole error bar, the integrated
uncertainties over the entire range leads to significant
changes. However, the agreement rapidly improves when adding more
modes, and with 3102 modes the agreement with the pixel basis is very
good. To quantify this statement, we calculate the integrated absolute
difference between the two distributions,
\begin{equation}
  \Delta = \int |\mathcal{L}_1 - \mathcal{L}_2|dqdn,
  \label{eq:qdef}
\end{equation}
and compare the resulting parameter with that computed from two
bi-variate Gaussians with identical covariances but different
means. Numerically, we find a value of $\Delta=0.002$ for the basis
including 3102 modes, which corresponds to a shift of $0.006\sigma$
for two bi-variate Gaussians. Recomputing the Fisher uncertainties with
this new basis, we find that the relative error increase is now
smaller than 3.8\% for all multipoles.

We emphasize that this particular number of modes is not to be taken as a
universal prescription, and in general will depend on the signal-to-noise
ratio of the data in question. However, we should note that the required
number of modes needed for convergence lie close to the number of modes
that are left after truncating the multipole expansion for temperature and
polarization at $\ell=32$, namely $3 \cdot (32 + 1)^2 = 3267$, with the
difference between this number and the required number of modes (3102) 
perhaps explained by the low signal-to-noise ratio of the polarization data.

\begin{figure}[t]
  \begin{center}
    \epsfig{figure=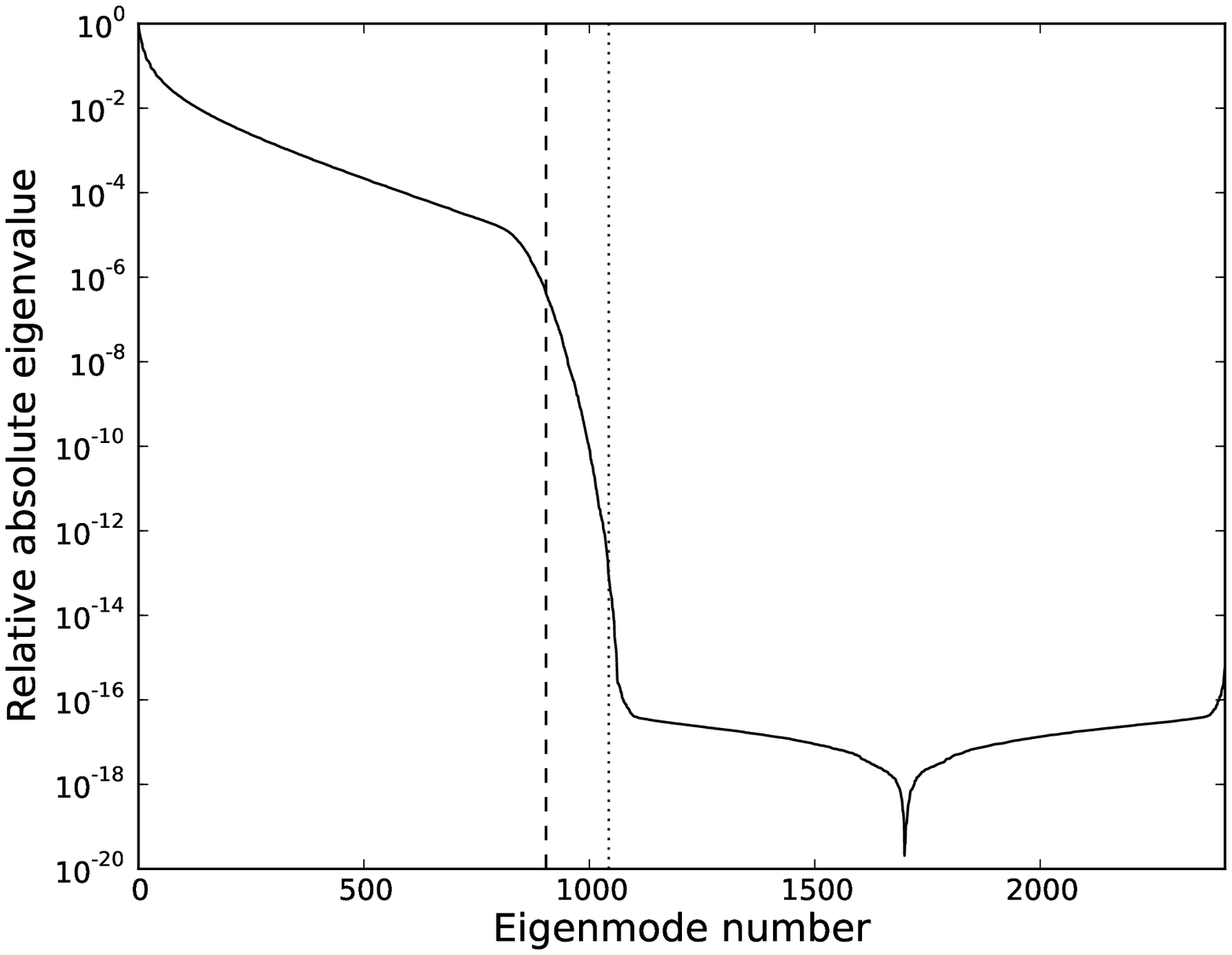, width=0.9\linewidth}
    \epsfig{figure=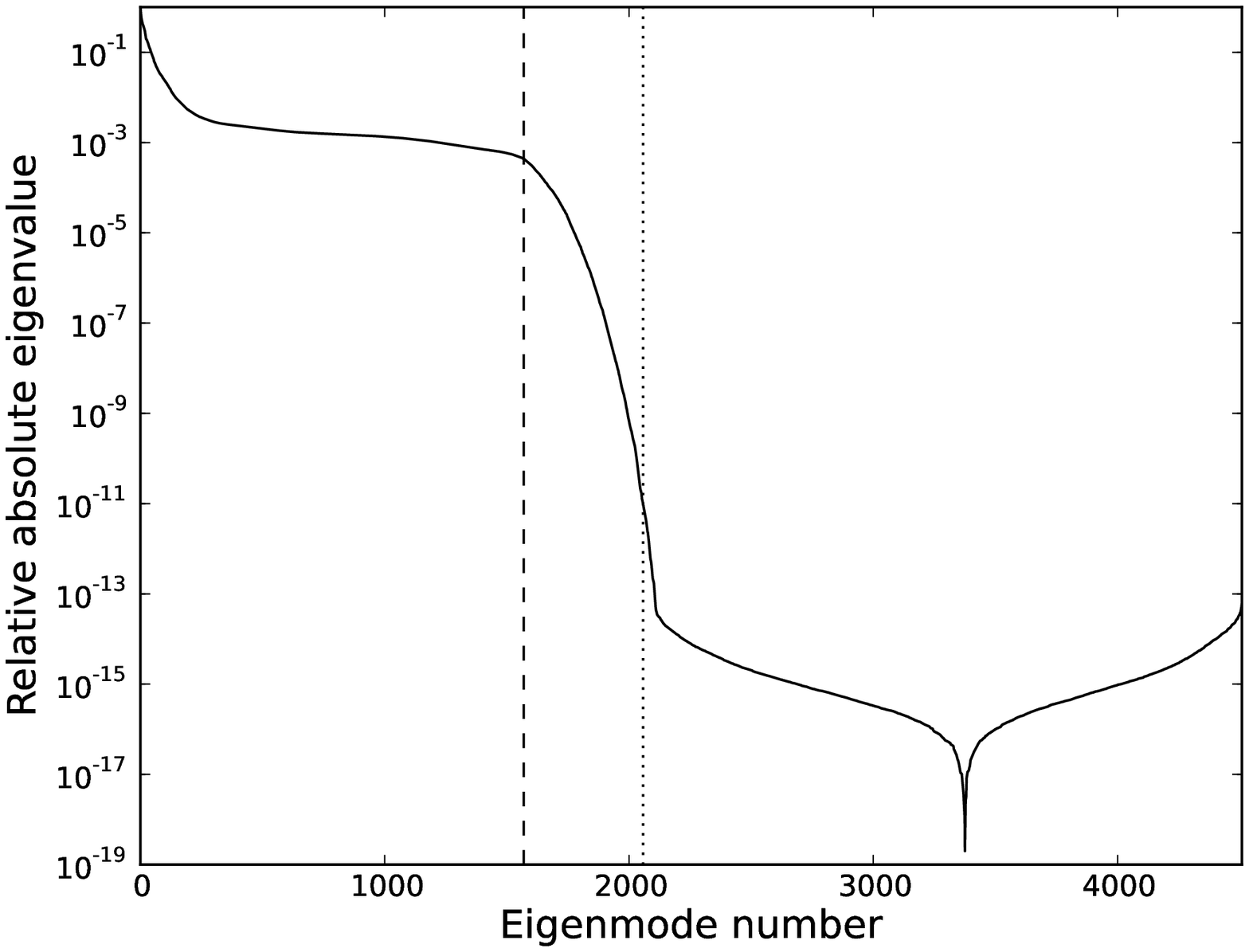, width=0.9\linewidth}
  \end{center}
  \caption{Normalized temperature (\emph{top}) and polarization
    (\emph{bottom}) eigenvalue spectrum for the signal-to-noise basis
    with $\ell_{\textrm{t}}=32$. The vertical lines indicate the
    cutoffs determined by the Fisher uncertainty (dashed) and
    amplitude--tilt (dotted) analyses. For robust results, all modes
    below the sharp decreases should be included.}
  \label{fig:eigenspectrum_ston}
\end{figure}

\begin{figure}[t]
  \begin{center}
    \epsfig{figure=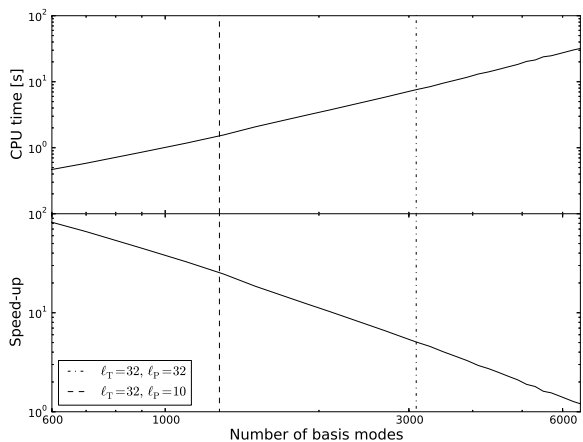, width=0.9\linewidth}
  \end{center}
  \caption{CPU time per likelihood evaluation as a function of number
    of basis modes (\emph{top}), and corresponding speed-up relative
    to pixel basis evaluation (\emph{bottom}). The average was
    calculated from 50 evaluations running on a single CPU. Vertical lines
    indicate the timing estimates for the two bases found in the text.}
  \label{fig:timing}
\end{figure}

Figure \ref{fig:eigenspectrum_ston} shows the eigenspectrum of the
signal-to-noise eigenbasis once again, this time with the two proposed
eigenmode cutoffs marked as vertical lines. To achieve reasonable
accuracy on individual multipoles, it is sufficient to include only
the high signal-to-noise modes in the flat high eigenvalue
plateaus. However, to properly account for correlations, it is
important to also include the modes that lie in the rapidly dropping
regime; these are partially degenerate modes that still carry some
information. On the other hand, beyond this rapid decrease the
remaining eigenvalues are for all practical purposes are zero, and can
be excluded safely. Note that this region starts around $(32+1)^2$ and
$2\cdot(32 + 1)^2$ for temperature and polarization, respectively,
which make intuitive sense, given that these are the number of modes
left after truncating the multipole expansion at $\ell=32$. Thus,
future basis optimization can be performed quite simply by computing
the the eigenspectrum of the signal-to-noise basis, and determining
the cutoff at which the numerically singular region begins. 

\begin{figure}[t]
  \begin{center}
    \epsfig{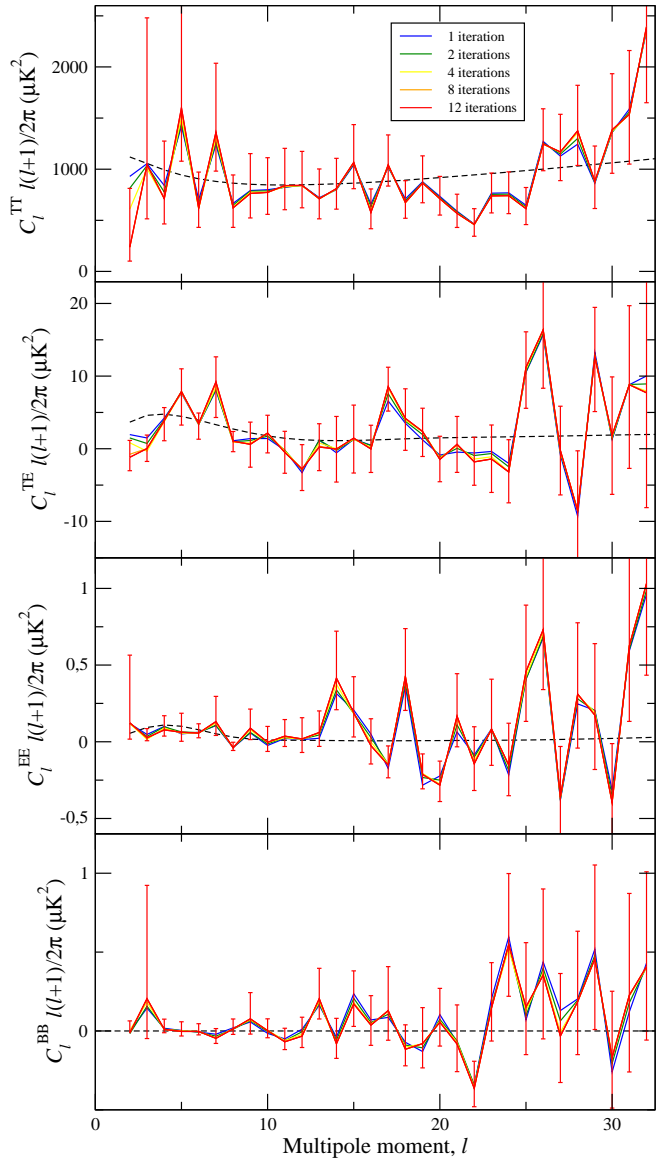}
  \end{center}
  \caption{9-year \WMAP\ QML power spectrum estimates as a function of
  QML iteration (from blue to red). Error bars are asymmetric 68\%
  confidence limits computed from the conditional likelihood evaluated
  around the maximum-likelihood point for the last iteration.}
  \label{fig:convergence}
\end{figure}

Before concluding this section, we show in Figure \ref{fig:timing} the
CPU time per likelihood evaluation as a function of number of basis
modes (top panel), as well as the corresponding speed-up (bottom
panel). Each point in this plot is computed as the average of 50
consecutive single-CPU evaluations. For the pixel basis that includes
6836 modes, each likelihood evaluation requires 35 CPU seconds, while
each signal-to-noise basis evaluation (including 3102 modes) requires
7.5 CPU seconds. The realized speed-up is thus a factor of 5, which,
although significant, is lower than the theoretical limit of
$(35/7.5)^3=11$ by a factor of two. On the other hand, all expensive
operations are implemented using standard Lapack routines, which are
already highly optimized, and fully gaining this factor is not
trivial.

\section{Power spectrum, likelihood and parameters}
\label{sec:results}

\begin{figure}[t]
  \begin{center}
    \epsfig{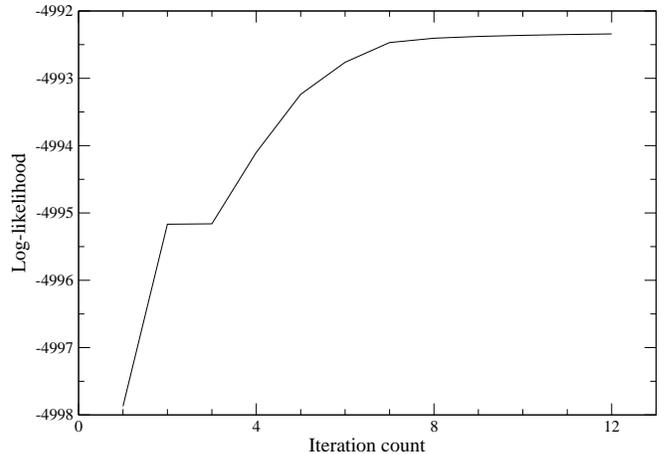}
  \end{center}
  \caption{Log-likelihood of the QML power spectrum estimate as a
    function of QML iteration.}
  \label{fig:lnL_qml}
\end{figure}

In this section we assess the performance of the compressed likelihood
formalism in terms of the CMB power spectrum, likelihood and
cosmological parameters. Figure \ref{fig:convergence} shows the
low-$\ell$ \WMAP\ temperature and polarization power spectrum as
derived with the QML estimator described in Section \ref{sec:qml},
using the signal-to-noise basis that contains 3102 basis modes from
the last section. Different colors show the result after different
number of QML iterations going from few (blue) to many (red). The
error bars indicate the asymmetric 68\% errors for the last
iteration. Figure \ref{fig:lnL_qml} shows the corresponding log-likelihood
as a function of iteration. 

Several interesting features may be seen in these plots. First of all,
the initial guess adopted for this calculation was the best-fit
\Planck\ 2013 model, indicated as dashed lines in Figure
\ref{fig:convergence}, while formal convergence was achieved after 12
iterations. However, we see that already a single QML iteration
results in a solution that for most multipoles is quite close to the
actual maximum-likelihood solution. For exploratory work, for instance
when trying to understand the effect of systematics on low-$\ell$
polarization studies, a single-iteration QML power spectrum
approximation may be quite useful, providing a near optimal power
spectrum estimate in less than 2 minutes of CPU time. However, for
final analysis it is clear that several iterations are indeed highly
desirable, as the log-likelihood increases by more than
$\Delta\ln\mathcal{L}=5.5$ and $\Delta\chi^2=-2\Delta\ln\mathcal{L}$
by more than 11. As a concrete example, the temperature quadrupole
converges slowly because of its intrinsically non-Gaussian shape, and
requires at least 8 iterations before stabilizing.

\begin{figure}[t]
  \begin{center}
    \epsfig{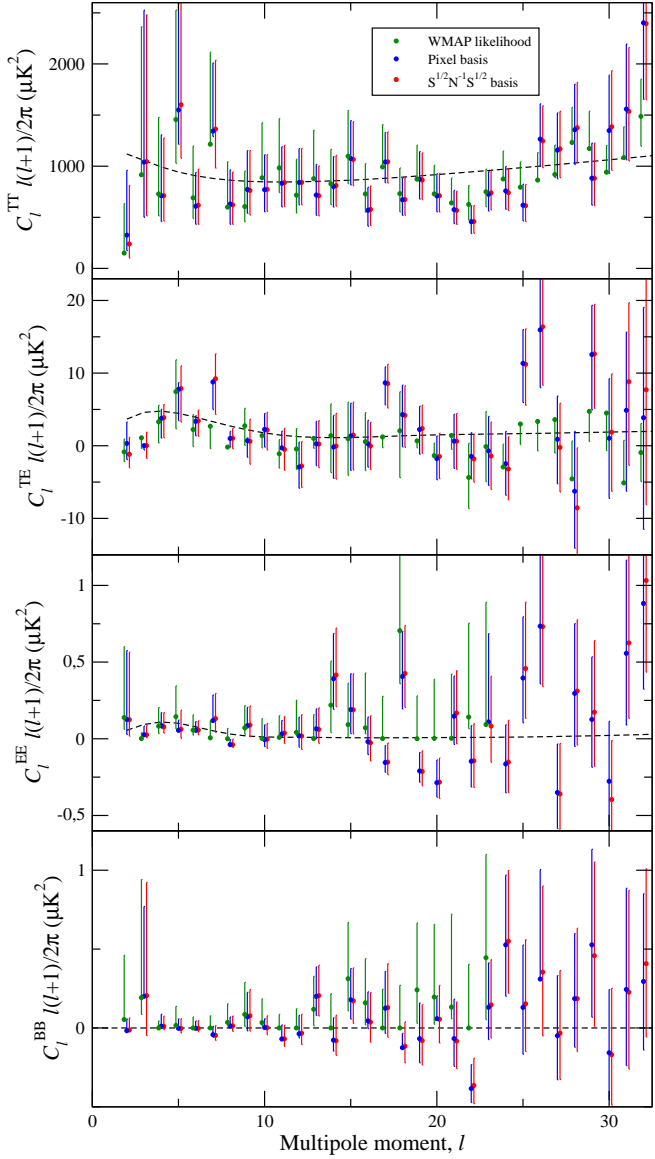}
  \end{center}
  \caption{Comparison of maximum-likelihood power spectra derived from
    three different \WMAP\ low-$\ell$ likelihood
    implementations. Error bars indicate asymmetric 68\% confidence
    regions computed from conditional likelihood slices around the
    joint maximum-likelihood point.}
  \label{fig:powspec}
\end{figure}

\begin{figure}[t]
  \begin{center}
    \epsfig{figure=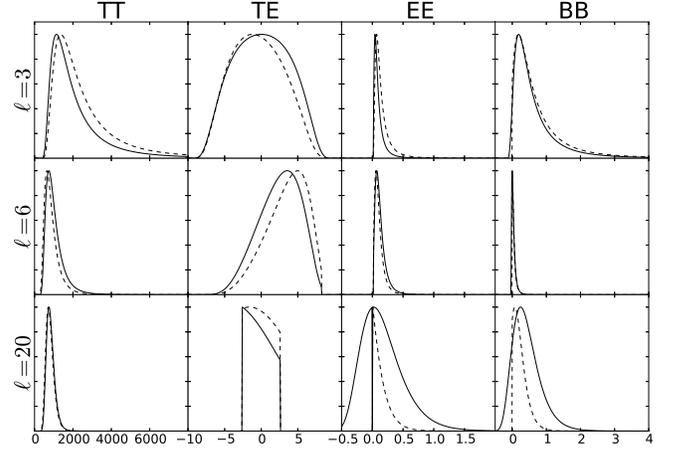, width=\linewidth}
  \end{center}
  \caption{Comparison of conditional likelihood slices computed from
    the official \WMAP\ likelihood (solid lines) and the
    signal-to-noise basis defined in the current paper (dashed
    lines). Other multipoles are fixed at the best-fit \Planck\ 2013
    $\Lambda$CDM power spectrum.}
  \label{fig:slices}
\end{figure}

In Figure \ref{fig:powspec} we compare three different power spectrum
estimates, all computed by maximum-likelihood techniques using the
9-year \WMAP\ data, but with different underlying likelihoods. Green
points are derived directly from the official \WMAP\ low-$\ell$
likelihood through a non-linear multivariate Powell search
\citep{press:2007}; blue points are derived from the pixel basis
likelihood described in this paper using the iterative QML estimator;
and red points are derived from the corresponding signal-to-noise
basis that includes 3102 modes. Figure \ref{fig:slices} compare
individual conditional likelihood slices computed from the \emph{WMAP}
likelihood and the signal-to-noise basis. 

\begin{figure}[t]
  \begin{center}
    \epsfig{figure=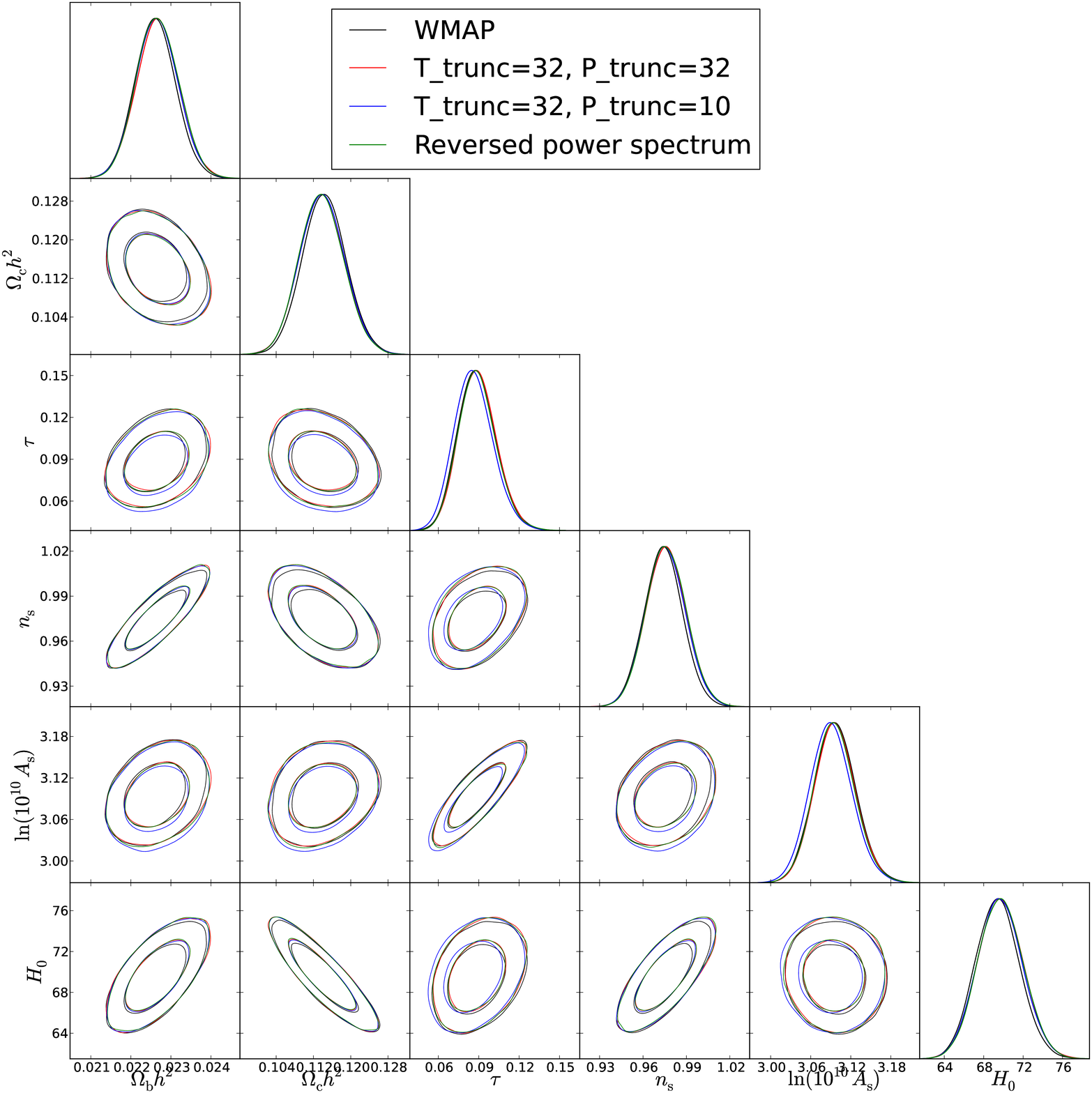, width=\linewidth}
  \end{center}
  \caption{Cosmological parameter distributions evaluated with CosmoMC
    for three different likelihoods; the official $WMAP$ likelihood
    (black), the signal-to-noise basis derived in the current paper,
    truncated at $\ell=32$ (red), the same but truncated $\ell=10$
    for polarization (blue), and the signal-to-noise basis using 
    a reversed power spectrum (green)}
  \label{fig:CosmoMC_plots}
\end{figure}

Overall, the agreement between the three spectra is very good, and for
most multipoles the relative shifts are much less than
$1\sigma$. However, there are notable differences as well, and perhaps
the most striking is the different behaviour of the error estimates
associated with the $C_{\ell}^{TE}$ spectrum. Considering first the
\WMAP\ spectrum in Figure \ref{fig:powspec}, a total of 16 out of 31
power spectrum coefficients have a vanishing error bar either toward
low of high values, indicating a maximum likelihood point that lies on
a sharp likelihood boundary. For comparison, for the pixel and
signal-to-noise basis likelihoods only 4 and 0 out of 31 coefficients
show similar behaviour, respectively. This difference is primarily due
to the different effective priors imposed by the two approaches; while
the \WMAP\ likelihood only requires the data covariance to be positive
definite, we impose the stronger criterion that the condition number
must also be well behaved (see Section \ref{sec:regprior}). The latter
prior prevents the nonlinear search algorithm from finding non-physical
power spectrum solutions near the singularity boundary with
artificially high likelihood values, which in turn forces correlated
power spectrum coefficients away from their best-fit values.

We also note that the pixel-based likelihood error bars are, in a sense,
less symmetric than those of the signal-to-noise basis. This is most likely
a cause of the fact that the pixel-based likelihood is more ill-conditioned
due to the higher number of redundant modes. This degeneracy makes the
likelihood slices behave worse for the pixel-based case than for the
signal-to-noise basis.

A different but related issue is seen in the plot of
$\mathcal{L}(C_{20}^{EE})$ in Figure \ref{fig:slices}. Here one can
see that the \emph{WMAP} likelihood allows significantly negative
values of $C_{20}^{EE}$, but not values very close to zero; there is a
``hole''in the likelihood surface. This is an artifact of the
temperature--polarization split implemented by the \WMAP\ likelihood,
in that positive definiteness is assessed separately for the
temperature and polarization components, effectively resulting in
three independent criteria (i.e., $C_{\ell}^{TT} > 0$ for the
Blackwell-Rao temperature component, $|\S+\N|>0$ for the polarization
component; and $C_{\ell}^{TE} < \sqrt{C_{\ell}^{TT}C_{\ell}^{EE}}$ for
the hybrid likelihood). With single joint likelihood implemented in
this paper, this problem becomes much simpler, in that there is only a
single (condition number based) numerical prior, and an optional
physical prior, $C_{\ell}^{TE} < \sqrt{C_{\ell}^{TT}C_{\ell}^{EE}}$,
whose valid parameter volume lies fully within the numerical prior
region.

%\subsection{Cosmological parameters}

%The last of these was used because it is assumed that the parameter $\tau$ is
%the only one affecting the $EE$ power spectra, and with the current best-fit
%values for $\tau$, all of the signal constraining $\tau$ will lie below
%$l=10$. The resulting number of eigenmodes found for this set is 1043
%temperature modes and 234 polarization modes - giving a significant (a factor of
%$\sim 2$) speedup for the complete CosmoMC run compared to the basis truncated
%at $\ell=32$ for both temperature and polarization.

%The results are shown in fig. \ref{fig:CosmoMC_plots}, and we summarize the
%parameter values in table \ref{tab:params}.

Our final test relates to cosmological parameters, as estimated using
CosmoMC \citep{lewis:2002} coupled to different versions of the 9-year
\WMAP\ likelihood. All cases used the same high-$\ell$ likelihood,
including $C_{\ell}^{TT}$ for $33 \le \ell \le 1200$ and
$C_{\ell}^{TE}$ for $33 \le \ell \le 800$, while at low $\ell$'s four
different variations were considered:
\begin{enumerate}
  \item The standard $WMAP$ low-$\ell$ hybrid
    temperature--polarization likelihood
  \item The signal-to-noise basis likelihood derived in Section
    \ref{sec:optimization} including 3102 modes.
  \item A similar signal-to-noise basis likelihood as (2), but with
    polarization truncated at $\ell=10$.
  \item The same as (2), but with the fiducial power spectrum used for
    basis definition extracted from an incorrect part of the original
    spectrum, specifically $D_{\ell} \leftarrow D_{1000-\ell}$
    with $D_{\ell}\equiv C_{\ell}\ell(\ell+1)/2\pi$.
\end{enumerate}
The latter is a simple test of potential sensitivity to the assumed
fiducial spectrum. 

\begin{deluxetable*}{lcccccccc}[t]
\tablewidth{0pt}
\tablecaption{\label{tab:params} Summary of cosmological parameters}
\tablecomments{Parameters derived with four different low-$\ell$
  \WMAP\ likelihood implementations; the official $WMAP$ low-$\ell$
  likelihood, and the signal-to-noise basis likelihoods derived in
  this paper, truncated at $\ell=32$ and 10, respectively, for
  polarization, and one truncated at $\ell=32$ with a reversed power
  spectrum as the basis signal.  The fourth, sixth, and eighth columns
  show the relative shifts with respect to the \emph{WMAP} approach
  measured in units of $\sigma$.  The confidence intervals are
  $1\sigma$, and the best-fit points are marginal posterior means.\vspace*{-5mm}
}
\tablehead{ & \textsc{Default WMAP}& \multicolumn{2}{c}{\textsc{P trunc = 32}} &
\multicolumn{2}{c}{\textsc{P trunc = 10}} & \multicolumn{2}{c}{\textsc{Reversed power spectrum}}\\
& Constraint & Constraint  & Deviation ($\sigma$) & Constraint  & Deviation ($\sigma$) & Constraint & Deviation ($\sigma$)}
\startdata
$\Omega_b h^2$              & $0.0226\pm 0.0005$ & $0.0227\pm 0.0005$ & $0.12$ & $0.0227\pm 0.0005$ & $0.09$ & $0.0227\pm 0.0005$ & $0.12 $\\
$\Omega_c h^2$              & $0.114 \pm 0.005 $ & $0.114 \pm 0.005 $ & $0.11$ & $0.114 \pm 0.005 $ & $0.09$ & $0.114 \pm 0.005$ & $0.13$ \\
$\theta$                    & $1.040 \pm 0.002 $ & $1.040 \pm 0.002 $ & $0.06$ &
$1.040 \pm 0.002 $ & $0.04$ & $1.040\pm 0.002$ & $0.05$ \\
$\tau$                      & $0.088 \pm 0.014 $ & $0.089 \pm 0.014 $ & $0.04$ &
$0.086 \pm 0.014 $ & $0.18$ & $0.089 \pm 0.014$ & $0.02$\\
$n_{s}$                     & $0.974 \pm 0.013 $ & $0.976 \pm 0.014 $ & $0.14$ &
$0.976 \pm 0.013 $ & $0.11$ & $0.976\pm 0.013$ & $0.15$ \\
$\textrm{log}[10^{10} A_s]$ & $3.10  \pm 0.03  $ & $3.10  \pm 0.03  $ & $0.001$
& $3.09  \pm 0.03  $ & $0.20$ & $3.10 \pm 0.03$ & $0.03$\\
$H_0$ & $69.4  \pm 2.17  $ & $69.7  \pm 2.22  $ & $0.13$ & $69.7  \pm 2.21  $ & $0.11$ & $69.7\pm 2.21$ & $0.14$
\enddata
\end{deluxetable*}

The results from these calculations are summarized in Figure
\ref{fig:CosmoMC_plots} and Table \ref{tab:params}. First and
foremost, we see that all results are highly robust against these
variations, with a maximum change of any marginal mean of at most
$0.2\sigma$. Of course, most of these parameters are dominated by
small-scale information, but also the optical depth of reionization,
$\tau$, which depends critically on the low-$\ell$ EE spectrum shows
very small variations. Even filtering away all polarization multipoles
above $\ell>10$ only affects the results by $0.18\sigma$. Finally,
reversing the power spectrum does not make any difference whatsoever,
with results that are identical to the default case up to the second
digit in the uncertainties.

\section{Summary}

Building on an idea proposed by \citet{tegmark:1997b}, we have
developed a framework for efficient low-$\ell$ CMB polarization
likelihood analysis using linear compression, and we have applied this
framework to the 9-year \WMAP\ data. Five different basis definitions
were compared in terms of compression efficiency, and, in agreement
with earlier suggestions, we find that an optimal basis may be defined
in terms of the eigenvectors of $\S^{1/2}\N^{-1}\S^{1/2}$, picking out
modes with high signal-to-noise. Within this basis, the original
low-$\ell$ \WMAP\ data set comprising 6834 pixels may be compressed
onto a smaller set of 3102 basis vectors with negligible loss of
accuracy, reducing the computational cost of a single likelihood
evaluation by a factor of five.

Next, we have used the same framework to implement an efficient and
stable version of the Quadratic Maximum Likelihood power spectrum
estimator, slightly re-writing the expressions for the covariance
matrix derivatives to use explicit projection operator. The
corresponding code requires about 3 GB of memory and 2 CPU minutes per
QML iteration for the \WMAP\ data at a HEALPix resolution of
$N_{\textrm{side}}=16$, which is well within the capabilities of a
standard laptop. Additionally, we have shown how to stabilize the QML
estimator, and avoid regions of parameter space in which the data
covariance matrix become non-positive definite. This increases the
overall computational cost by a small factor, as it relies on a
non-linear optimization within each main QML iteration, but for most
cases this is not a major problem. 

On a related topic, we have introduced a new and more effective prior
for removing nonphysical artifacts on the likelihood
surface. Previously, this was done by only requiring the data
covariance matrix to be positive definite, but this leaves significant
anomalies near the singularity boundary. A better option is to put a
constraint on the condition number of the covariance matrix. 

Finally, we note that while only the \WMAP\ likelihood was considered
in this paper, we expect that the methods presented here should be
directly applicable to the up-coming \Planck\ polarization
data.

\begin{acknowledgements}
  We would like to thank Antony Lewis for useful discussions. This
  project was supported by the ERC Starting Grant StG2010-257080.
  Part of the research was carried out at the Jet Propulsion
  Laboratory, California Institute of Technology, under a contract
  with NASA.  Some of the results in this paper have been derived
  using the HEALPix \citep{gorski:2005} software and analysis package.
\end{acknowledgements}

\end{document}